\documentclass[conference]{IEEEtran}
\IEEEoverridecommandlockouts
\makeatletter
\def\ps@headings{%
\def\@oddhead{\mbox{}\scriptsize\rightmark \hfil \thepage}%
\def\@evenhead{\scriptsize\thepage \hfil \leftmark\mbox{}}%
\def\@oddfoot{}%
\def\@evenfoot{}}
\makeatother

\usepackage{amssymb,color,amsmath}
\usepackage{paralist}
\usepackage{graphicx}

\newcommand{\F}{\mathbf{F}}

\newcommand{\C}{\mathcal{C}}
\newcommand{\N}{\mathcal{N}}
\newcommand{\G}{\mathcal{G}}
\newtheorem{theorem}{\textbf{Theorem}}
\newtheorem{lemma}[theorem]{\textbf{Lemma}}
\newtheorem{definition}[theorem]{\textbf{Definition}}

\newtheorem{example}[theorem]{\textbf{Example}}

\newcommand{\nix}[1]{}

\begin{document}
\title{Network Protection Codes Against Link Failures Using Network Coding\thanks{This
research was supported in part by grants CNS-0626741 and
CNS-0721453 from the National Science Foundation,
and a gift from Cisco Systems.}}
\author{
Salah A. Aly \hspace*{.3in} and \hspace*{.3in}
Ahmed E. Kamal\\
Department of Electrical and Computer Engineering\\
Iowa State University, Ames, IA 50011, USA\\Email: \{salah,kamal\}@iastate.edu}

\maketitle

\begin{abstract}
Protecting against link failures in communication networks is
essential to increase robustness, accessibility, and reliability of
data transmission. Recently, network coding has been proposed as a
solution to provide agile and cost efficient network protection
against link  failures, which does not require data rerouting, or
packet retransmission. To achieve this, separate paths have to be
provisioned to carry encoded packets, hence requiring either the
addition of extra links, or reserving some of the resources for this
purpose.  In this paper, we propose  network
protection codes against a single link failure using network coding, where
a separate path using reserved links is not needed.  In this case
portions of the link capacities are used to carry the encoded
packets.

 The scheme is  extended to protect against multiple
link failures and can be implemented  at an overlay layer. Although
this leads to reducing the network capacity, the network capacity
reduction is asymptotically small in most cases of practical
interest. We demonstrate that such \emph{network protection codes} are equivalent to error correcting codes for erasure channels. Finally, we study the encoding and decoding operations of such codes over the binary field.
\end{abstract}

\section{Introduction}\label{sec:intro}
Network coding is  a powerful tool that has been used to increase
the throughput, capacity, and performance of communication
networks~\cite{soljanin07,yeung06}. It offers benefits in terms of
energy efficiency, additional security, and reduced delay. Network
coding allows the intermediate nodes not only to forward packets
using network scheduling algorithms, but also encode/decode them
using algebraic primitive operations
(see~\cite{ahlswede00,fragouli06,soljanin07,yeung06} and references
therein).

One application of network coding that has been proposed recently is
to provide protection against link failures in overlay
networks~\cite{kamal06a,kamal08b}. This is achieved by transmitting
combinations of data units from multiple connections on a backup
path in a manner that enables each receiver node to recover a copy
of the data transmitted on the working path in case the working path
fails. This can result in recovery from failures without data
rerouting, hence achieving agile protection. Moreover, the sharing
of protection resources between multiple connections through the
transmission of linear combinations of data units results in
efficient use of protection resources. This, however, requires the
establishment of extra paths over which the combined data units are
transmitted. Such paths may require the addition of links to the
network under the Separate Capacity Provisioning strategy (SCP), or
that paths be provisioned using existing links if using the Joint
Capacity Provisioning strategy (JCP), hence reducing the network
traffic carrying capacity.

Certain networks can allow extra transmissions and the addition of
bandwidth, but they do not allow the addition of new paths.   In this scenario,
one needs to design efficient data recovery schemes. Several
previous approaches focused on solving this problem using additional
extra paths at an overlay network level, or deploying ARQ protocols
for the recovery of lost packets.  In order to provide recovery from
link failures in such networks, approaches other than using
dedicated paths, or adding extra links must be used. In this paper,
we propose such an approach in which we use network coding to
provide agile, and resource efficient protection against link
failures, and without adding extra paths. The approach is based on
combining data units from a number of sources, and then transmitting
the encoded data units using a small fraction of the bandwidth
allocated to the connections, hence disposing of the requirement of
having extra paths. In this scenario, once a path fails, the
receiver can recover the lost packets easily from the neighbors by
initiating simple queries.

\goodbreak
Previous solutions in network survivability approaches using network
coding focused on providing backup paths to recover the data
affected by the
failures~\cite{kamal06a,kamal07a,kamal07b}. Such
approaches include 1+N, and M+N protections. In 1+N protection, an
extra secondary path is used to carry combinations of data units
from N different connections, and is therefore used to protect N
primary paths from any single link failure. The M+N is an extension
of 1+N protection where M extra secondary paths are needed to
protect multiple link failures.

\goodbreak

In this paper, we apply network coding for network protection
against link failures and packet loss. We define the concept of
protection codes similar to error-correcting codes that are widely
used in channel coding~\cite{huffman03,macwilliams77}. Protection
codes are a new class of error monitoring codes that we propose in
Section~\ref{sec:multiplefailures}. Such codes aim to provide better
provisioning and data recovery mechanisms. A protection code is
defined by a matrix $G$ known at a set of senders S and  receivers
R. Every column vector in the generator matrix of a protection code
defines the set of operations, in which every sender (receiver)
needs to perform. \bigbreak

\goodbreak
The new contributions in this paper are stated as follows:
\begin{compactenum}[i)]
\item We introduce link protection  network coding-based using reduced capacity instead of
adding extra paths as shown in the previous
work~\cite{kamal06a,kamal07a,kamal07b}.

\item
We develop a theoretical foundation of protection codes, in which
the receivers are able to recover data sent over $t$ failed links
out of $n$ primary links.

\end{compactenum}


This paper is organized as follows. In Section~\ref{sec:relatedwork}
we briefly state the related work and previous solutions to the
network protection problem using network coding.  In Section~\ref{sec:networkmodel} we
present the network model and problem definition.
Sections~\ref{sec:singlefailure} and~\ref{sec:multiplefailures}
discuss single and multiple link failures  and how to protect these
link failures using reduced capacity and network coding. In
Section~\ref{sec:analysis} we give analysis of the general case of
$t \ll n$ link failures, and the paper is concluded in
Section~\ref{sec:conclusion}.

\section{Related Work}\label{sec:relatedwork}

In~\cite{kamal06a}, the author introduced a 1+N protection model in
optical mesh networks using network coding over p-cycles. The author
suggested a model for protecting $N$ connections  from a set of
sources to a set of receivers in a network with $n$ connections,
where one connection might fail. The suggested model can protect
against a single link failure in any arbitrary path connecting a
source and destination.

In~\cite{kamal07a}, the author extended the previous model to
protect multiple link failures. It is shown that protecting against
$m$ failures, at least $m$ p-cycles are needed. An illustrative
example in case of two link failures was given. The idea was to
derive $m$ linearly independent equations to recover the data sent
from $m$ sources.

In~\cite{kamal07b}, the author extended the protection model
in~\cite{kamal06a} and provided a GMPLS-based implementation of a
link protection strategy that is a hybrid of 1+N and 1:N. It is
claimed that the hybrid 1+N link protection provides protection at
higher layers and with a speed that is comparable to the speed
achieved by the physical layer implementations. In addition, it has
less cost and much flexibility.

 Monitoring
network information flow using network coding was introduced
in~\cite{ho05,ho03}. In~\cite{fragouli05}, it was shown how to use
network coding techniques to improve network monitoring in overlay
networks. Practical aspects of network coding has been shown
in~\cite{chou03}.

In this paper, we provide a new technique for protecting network
failures using \emph{protection codes} and \emph{reduced capacity}.
This technique can be deployed at an overlay layer in optical mesh
networks, in which detecting failure is an essential task. The
benefits of the proposed  approach are that:
\begin{compactenum}[i)]
\item
It allows receivers to recover the lost data without {data
rerouting or data retransmission}.
\item
It has less computational complexity and does not require adding
extra paths {or reserving backup paths}.
\item
{At any point in time,}
all $n$ connection paths have full capacity except at one path in case of
protecting against a single link failure and $m < n$ paths in case
of protecting against $ m $ link failures.
\end{compactenum}

We will analyze the proposed \emph{ protection codes} and error correcting codes  that are used for erasure channels.

\section{Network Model}\label{sec:networkmodel}

Let $\mathcal{G}=(V,E)$ be a graph which represents an abstraction
of a  set of connections.  $V$ is a set of network nodes and $E$ is
a set of edges. Let there be $n$ unicast connections,
and let $S \subset V$ be the set of sources $\{s_1,...,s_n
\}$ and $R \subset V\backslash S$ be the set of receiver nodes $\{
r_1,...,r_{n} \}$ of the $n$ connections in $ \mathcal{G}$. The case of
$S \cap R \neq \phi$ can be easily incorporated in our model. Two nodes $u$  and $v$ in $ V \backslash \{S\cup R\}$  are connected by an edge $(u,v)$ in $E$ if there is a
direct connection between them.  We assume that the sources are
independent of each other, meaning they can only send messages and
there is no correlation between them. For simplicity, we will assume
that a direct disjoint path exists between $s_i$
and $r_i$, {and it is disjoint from the path between $s_j$ and $r_j$, for $j \neq i$}.

The graph $\G$ represents an abstraction of our network model $\N$
with the following assumptions.

\begin{compactenum}[i)]
\item  Let
$\N$ be a network with a set of sources $S=\{s_1,s_2,\ldots,s_n\}$
and a set of receivers $R=\{r_1,r_2,\ldots,r_n\}$, where $S \cup
R \subset V$.

\item Let $L$ be a set of links $L_1,L_2,\ldots,L_n$ such that there
is a link $L_i$ if and only if there is a connection path between
the sender $s_i$ and receiver $r_i$, i.e.,
\begin{eqnarray} L_i=
    \{(s_i,w_{1i}),(w_{1i},w_{2i}),\ldots,(w_{(m)i},r_i) \},\end{eqnarray} where
    $1\leq i\leq n$ and $(w_{(j-1)i},w_{ji}) \in E$, for some integer m.
Hence
we have $|S|=|R|=|L|=n$. The n connection paths are pairwise link
disjoint.

\item Every source $s_\ell$ sends a packet with its own $ID_{s_\ell}$ and data
 $x_\ell$ to the receiver $r_\ell$, so
 \begin{eqnarray}
packet_{s_\ell}=(ID_{s_\ell},x_\ell, t_\ell^\delta),
 \end{eqnarray}
where $t_\ell^\delta$ is the round time at step $\delta$ of the
source packet $packet_{s_\ell}$.
\item All links carry uni-directional messages from sources to
receivers.
\item  We consider the scenario where the cost
of adding a new path is higher than just combining messages in an
existing path, or there is not enough resources to provision extra
paths in the network. These two cases correspond to separate and
joint capacity provisioning, respectively~\cite{zhou00}.
\end{compactenum}

We can define the unit capacity $c_i$ of a link $L_i$ as follows.

\medskip

\begin{definition}
Let $\N$ be a network model defined by a tuple $(S,R,L)$. The unit
capacity of a link $L_i$ is given by
\begin{eqnarray}
 c_i= \left\{
  \begin{array}{ll}
    1, & \hbox{ $L_i$ is active;} \\
    0, & \hbox{otherwise .}
  \end{array}
\right.
\end{eqnarray}
Also, the average normalized capacity of $\N$ is defined by the
total number of active links divided by the total number of links
$n$
\begin{eqnarray}
C_{\N}=\frac{1}{n}\sum_{i=1}^n c_i.\end{eqnarray}
\end{definition}

This means that each source $s_i$  can send one packet per unit time
on a link $L_i$. Assume that all links have the same capacity. In
fact, we measure the capacity of $\N$ in the sense of the max-flow
min-cut theorem, see~\cite{karger99}. One can always assume that a
source with a large rate can be divided into a set of sources, each
of which has a unit link capacity.
\goodbreak

We can also define the set of sources that are connected to a source
$s_i$ in $\N$ as the degree of this source.

\medskip

\begin{definition}\label{def:nodedegree}
The number of neighbors with a direct connection to a node $u$
(i.e., a source $s_i$ in $S$ in the network $\N$) is called the
\emph{node degree} of $u \in V$, and is denoted by $d_n(u)$, i.e.,
\begin{eqnarray}
1\leq |\N(u)|=d_n(u)\leq n.
\end{eqnarray}
\end{definition}

The following definition describes the \emph{working} and
\emph{protection} paths between two network components.

\begin{definition}
The\emph{ working paths} on a network with n connection paths carry
traffic under normal operations. The \emph{Protection paths} provide
alternate backup paths to carry the traffic in case of failures. A
protection scheme ensures that data sent from the sources will reach the
receivers in case of failure incidences on the working paths.
\end{definition}

In this work the  goal is to provide  a reliable method for data protection sent
over a link $L_i$ without adding extra paths to the existing ones,
but by possibly reducing the source rates slightly. In fact there
are network scenarios where adding extra path is not
applicable~\cite{somani06,vasseur04,zhou00}. We propose a model to
protect link failures using network coding where some senders are
able to encode other sender's packets. We will study the network
protection against link failures at an overlay layer in two cases:
Single link failures and multiple link failures.

\section{Protecting Networks Against  A Single Link Failure}\label{sec:singlefailure}
In this section we study the problem of protecting a set of
connections against a single link failure in a network $\N$ with a
set of sources $S$ and a set of receivers $R$. This problem has been
studied in~\cite{kamal06a,kamal07a} by provisioning a path that is
link disjoint from all connection paths, and passes through all
sources and destinations.  All source packets are encoded in one
single packet and transmitted over this path. The encoding is
dynamic in the sense that packets are added and removed at each
source and destination.

Assume  that the  assumptions about the proposed network model $\N$, and the
abstraction graph $\G$ presented in Section~\ref{sec:networkmodel}
hold. We know that if there is an active link $L_i$ between $s_i$
and $r_i$, then the capacity $c_i$ is the unit capacity.
{Let us
consider the case where every source $s_i$ sends its own data $x_i$ and the encoded
data $y_i$}.
The encoded  message $y_i$
is defined as \begin{eqnarray}y_i=x_1\oplus\ldots \oplus x_{i \neq j} \oplus\ldots \oplus x_n\end{eqnarray} from all
other sources $S\backslash \{s_i\}$ over the finite field
$\F_2=\{0,1\}$, where the symbol $\oplus$ is the XOR operation.

Assume that among the set of links $L$, there is a  link $L_i$
for $1 \leq i  \leq n$ such that the sources $s_i$  sends a packet to the receivers $r_i$
as follows

 \begin{eqnarray}
packet_{s_i}=(ID_{s_i},x_i,
t_i^\delta).
 \end{eqnarray}

Assume for now that link $L_j$ has
the unit capacity.  The source $s_j$ sends a packet that will carry the encoded data $y_j$ to the receiver $r_j$  over the link $L_j$,
 \begin{eqnarray}
packet_{s_j}=(ID_{s_j},y_j, t_j^\delta).
 \end{eqnarray}
We assume that the summation operations are performed over $\F_2$.

Now we consider the case where
there is a single failure in a link $L_k$. Therefore, we have two
cases:
\begin{compactenum}[i)]
\item
If $k \neq j$, then the receiver $r_k$ needs to query $(n-1)$
nodes in order to recover the lost data $x_k$ over the failed link
$L_k$.
\typeout{The reason is that $x_k$ exists either at  $r_j$, and
it requires information of all other receivers. Hence, the lost data is a plain
message.}
{$x_k$ can be recovered by adding all other $n-1$ data units.}

\item
If the link $L_j$  has a failure, then
the receiver $r_j$ does not need to query any other node. In this case the link $L_j$ carries encoded data that is used for protection.
\end{compactenum}

This shows that only one single receiver needs to perform $(n-2)$
operations in order to recover its data if its link fails. In
other words, all other receivers will receive the transmitted data
from the senders of their own connections with a constant operation
$O(1)$.

\subsection{Network Protection Codes (NPC) for a Single Link Failure}
We can define the set of sources that will send encoded packets by
using constraint matrices.  We assume that there is a network
protection code $\C \subseteq \F_2^{n}$ defined by the constraint systematic  matrix

\begin{eqnarray}\label{eq:G}
G\!\!=\!\! \left[ \begin{array}{ccccccccc}1&0&\ldots&0&1\\ 0&1& \ldots&0&1\\
\vdots&\vdots&\vdots&\vdots&\vdots\\
 0&0&\ldots&1&1\end{array}\right]_{(n-1) \times n} \!\!,
\end{eqnarray}

\begin{figure}[t]
 \begin{center} 
  \includegraphics[scale=0.65]{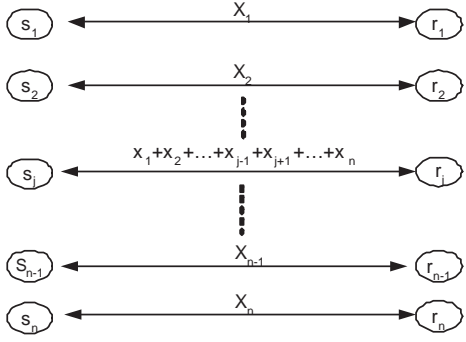}
  \caption{Network protection against a single link failure using reduced capacity and network coding. One link out of  $n$ primary links  carries encoded data.}\label{fig:nnodes}
\end{center}
\end{figure}

Without loss of generality, in Equation~(\ref{eq:G}), the  column vector $(\begin{array}{ccccc}
g_{1j}&g_{2j}&\ldots&g_{(n-1)j}\end{array})^T$ in $\F_2^{n-1}$ corresponds
to (n-1) sources, say for example the sources $s_1,s_2,\ldots,s_{n-1}$, that will send (update) their
values to (n-1) receivers, say i.e., $r_1,r_2,\ldots,r_{n-1}$. Also, there exists one source that will send encoded data.  Also, the row vector
$(\begin{array}{ccccc} g_{i1}&g_{i2}&\ldots&g_{in}\end{array})$ in
$\F_2^n$ determines  the channels $L_1,L_2,\ldots,L_n$. The column vector $g_{in}$  corresponds
to the source $s_i$ that will carry encoded data on the
connection path $L_i$, see
Fig.~\ref{fig:nnodes}. The minimum weight of a row in $G$ is 2.

\smallskip

We can define the \emph{protection codes} that will protect a single path failure as follows:
\begin{definition}
An $[n,n-1,2]_2$  \emph{network protection code} $\C$ is a $n-1$~dimensional
subspace of the space $\F_2^{n}$ defined by the systematic generator matrix
$G$ and is able to protect a single network failure of an arbitrary
path  $L_i$.
\end{definition}

We note that the \emph{protection codes} are also error correcting codes that can be used for channel  detection. Recall that an $[n,n-1,2]$ code over $\F_2$ is a code that encodes (n-1) symbols into n symbols and detects (correct from) a single path failure.

In general, we will assume that the code $\C$ defined by the
systematic generator matrix $G$ is known for every source $s_i$ and every
receiver $r_i$. This means that every receiver will be able to
recover the data $x_i$ if the link $L_i$ is corrupted. We assume that the positions of the failures are known. Furthermore,
every source node has a copy of the code $\C$. Without loss of generality, the protection matrix
among all sources is given by:

\begin{eqnarray}\label{eq:matrixXi}
\begin{array}{c||cccccc}
&L_1&L_2&\cdots&L_{n-1} &L_n&\\ \hline \hline
s_1&x_1&0&\cdots&0&x_{1}\\
s_2&0&x_2&\cdots&0&x_{2}\\
\vdots&\vdots&\vdots&\cdots&\vdots&\vdots\\
s_{n-1}&0&0&\cdots&x_{n-1}&x_{n-1}\\
\hline \hline total&x_1&x_2&\ldots&x_n&y_n\\
\end{array}
\end{eqnarray}

 $y_n$ is the protection value from   every source $s_\ell$
that will be encoded at source $s_n$, for
all $1\leq \ell \leq n-1$.  Put differently, we have
\begin{eqnarray}\label{eq:protectionvalues}
y_n=\sum_{\ell=1}^{n-1}  x_{\ell}
\end{eqnarray}
The summation operation is defined by the XOR operation.  We note that the any source $s_i$ can carry the encoded data.

Hence from the matrix (\ref{eq:matrixXi}), we have
\begin{eqnarray}\label{eq:protectionvalues2}
y_{j}=\sum_{\ell=1, \ell \neq j}^n x_{\ell}
\end{eqnarray}

We assume that every source $s_j$ has a buffer that   stores its
value $x_j$ and the protection value $y_{j}$. Hence $s_j$ prepares
a packet $packet_{s_j}$ that contains the values
\begin{eqnarray}
packet_{s_j}= (ID_{s_j}, y_{j}, t_\ell^\delta),
\end{eqnarray}
where $y_{j}$ is defined in Equation~(\ref{eq:protectionvalues2}).

\begin{example}
Let $S$ and $R$ be two sets of senders and receivers, respectively, in the network model $\N$. The following scheme explains the plain and encoded data sent in five consecutive rounds from the five senders to the five receivers.
\begin{eqnarray}
\begin{array}{|c|ccccc|c|c|}
\hline
cycle& &~~~~1&&&&2&3\\
\hline
rounds&1&2&3&4&5&\ldots&\ldots\\
\hline
\hline
s_1 \rightarrow r_1 &y_1&x_1^1  &x_1^2 &x_1^3 &x_1^4& \ldots&\ldots\\
s_2 \rightarrow r_2 &x_2^1&y_2  &x_2^2 &x_2^3&x_2^4&\ldots&\ldots\\
s_3 \rightarrow r_3 &x_3^1  &x_3^2&y_3 &x_3^3&x_3^4&\ldots&\ldots \\
s_4 \rightarrow r_4     &x_4^1 &x_4^2 & x_4^3&y_4&x_4^4&\ldots&\ldots\\
s_5 \rightarrow r_5     &x_5^1 &x_5^2 & x_5^3&x_5^4&y_5&\ldots&\ldots\\
\hline
\end{array}
\end{eqnarray}
The encoded data $y_j$, for $1 \leq j \leq 5$,  is sent as

\begin{eqnarray}
y_j=\sum_{i=1}^{j-1} x_i^{j-1}+\sum_{i=j+1}^5 x_i^{j}.
\end{eqnarray}
We notice that every message has its own round. Hence the protection data is distributed among all paths for fairness, see~\cite{aly08j} for further details.
\end{example}

\nix{
\begin{example}
Consider five sources $\{s_1,s_2,s_3,s_4,s_5\}$ and five receivers
$\{r_1,r_2,r_3,r_4,r_5\}$. Without
loss of generality, let us assume that the source $s_i$ sends its
message $x_i$ to the receiver $r_i$ for $i=\{1,2,3,4\}$. Furthermore,
the source $s_5$ sends the message
$x_1\oplus x_2 \oplus x_3\oplus x_4$ to the receiver $r_5$.
This is an example where a single path failure can be recovered
from using network coding and the protection code shown above.

Hence, the source $s_5$   prepares the message
$y_1=x_1\oplus x_2\oplus x_3\oplus x_4$, and sends the packet
$$packet_{s_5}=(ID_{s_5},y_1, t_{5}^\delta).$$
Also, for $i=\{1,2,3,4\}$, the source $s_i$ sends the packet
$$packet_{s_i}=(ID_{s_i},x_i, t_{i}^\delta).$$

 So, every receiver $r_\ell$ will
obtain  a packet at a round time $t_\ell^\delta$ in a connection
path $L_\ell$. If we assume that there is one failed path, then four
receivers will receive their packets correctly. Assuming a receiver,
with a failure in its path, knows the matrix $G$, in this case it is
able to query other receivers to obtain its data.
\end{example}
}

 \smallbreak

We notice that it is  enough to allow only one source node to
perform the encoding operations for protecting against a single path
failure. This fact can be stated in the following lemma.


\begin{lemma}
Encoding the data from sources $S\backslash \{s_i\}$ at a source
$s_i$ in the network $\N$ is  enough to protect against a single
path failure.
\end{lemma}


\begin{lemma}
The total number of encoding operations needed to recover from a
single link failure in a network $\N$ with $n$ sources is given by
 {$(n-2)$} and the total number of transmissions  is $n$.
\end{lemma}

The previous lemma guarantees the recovery from  a single arbitrary
link failure. The reason is that the link that carries encoded data
might fail itself and one needs to protect its data.

\medbreak

\begin{lemma}\label{lem:capacitysinglelink}
In  the network model $\N$, the average network capacity of
protecting against a single link failure using reduced capacity and
network coding is given by $(n-1)/n$.
\end{lemma}
\begin{proof}(Sketch)

\begin{inparaenum}[i)] \item We know that every source $s_\ell$ that sends the data
$x_\ell$ has capacity $c_{\ell}=1$. \item Also, the source $s_i$
that sends $x_i$ and the encoded data $y_{i}$ at  different slots,
has a full capacity.
\item The source $s_i$ is not fixed among all nodes $S$,
however, it is  rotated periodically over all sources for
fairness. On average one source of the $n$ nodes will reduce its
capacity. This shows the capacity of $\N$ as stated.
\end{inparaenum}
\end{proof}

\section{Protecting Networks Against Multiple
Link Failures}\label{sec:multiplefailures}

In the previous section we introduced a strategy for a single link failure in optical mesh
networks, where the chance of a single link failure is much higher
than multiple link failures. However, it was shown
in~\cite{fragouli05} through an experimental study that about
$\%30$ of the failures of the Sprint backbone network are multiple
link failures. Hence, one needs to design a general strategy against
multiple link failures.

In this section we will generalize the above strategy
to protect against $t$ path
failures using \emph{network protection codes} (NPC) and the reduced
capacity. We have the following assumptions about the channel model:
\begin{compactenum}[i)]
\item We assume that any $t$ arbitrary paths may fail and they
are independent of each other.
\item Location of the failures are known, but they are arbitrary among n connections.
    \item
    Protecting n working paths, k connection must carry plain data, and $m=n-k$ connections must carry encoded data.
\item We do not add extra
link paths, and every source  node is able to encode  the incoming
packets.
\item We consider the encoding and decoding operations are performed over $\F_2$.
\end{compactenum}
We will show the connection between error correcting
codes and protection codes~\cite{huffman03,macwilliams77}.

We have $n$ working paths from the senders to receivers. We will
assume that a  path $L_i$ can have a full capacity or it can manage
a buffer that maintains the full capacity where the encoded data is sent.

Assume that the notations in the previous sections hold. Let us
assume a network model $\N$ with $t>1$ path failures.  One can
define a protection code $\C$ which protects $n$ links as shown in
the systematic  matrix $G$ in~(\ref{eq:Gmultiple}). In general, the systematic generator matrix $G$
defines the source nodes that will send encoded  messages and source
nodes that will send only plain messages. In order to protect n working paths, k connection must carry plain data, and $m=n-k$ connections must carry encoded data. The
systematic generator matrix of the NPC for multiple link failures is given by:


\begin{eqnarray}\label{eq:Gmultiple}
G= \left[\!\!\!
\begin{array}{c|c}
\begin{array}{cccc}1&0&\ldots&0\!\!\!\\
0& 1&\ldots&0\!\!\!\\
\vdots&\vdots&\vdots&\!\!\!\\
 0&0&\ldots&1\!\!\!\\
 \end{array}\!\!\!&\begin{array}{ccccc}p_{11}\!&\!p_{12}&\ldots&p_{1m}\\
p_{21}\!&\!p_{22}&\ldots&p_{2m}\\
\vdots&\vdots&\vdots&\vdots\\
p_{k1}&p_{k2}&\ldots&p_{km}\\
 \end{array} \\\\
 \multicolumn{2}{c}{}\\ \underbrace{\hskip 0.7in}^{\mbox{identity matrix $I_{k\times k}$}} &\underbrace{\hspace{0.7in}}^{\mbox{ Submatrix } P_{k \times m}}\\
\end{array}\!\!\!
  \right]\!,\!\!\!\!
\end{eqnarray}
where $p_{ij} \in \F_2$
\goodbreak

The matrix $G$ can be rewritten as
\begin{eqnarray}
G= \big[\mbox{ } I_k \mbox{ } \mid  \mbox{ } \textbf{P} \mbox{ }
\big],
\end{eqnarray}
where $\textbf{P}$ is the sub-matrix that defines the redundant data
$\sum_{i=1}^k p_{ij}$, for $1 \leq j  \leq m$, to be sent to a set of sources for the purpose
of data protection against  data loss and link protection against
link failures. Based on the above matrix, every source $s_i$ sends
its own message $x_i$ to the receiver $r_i$ via the link $L_i$. In
addition $m$ links out of the $n$ links will carry encoded data. $d_{min}$ is the minimum weight of a row in $G$.

\smallskip

\begin{definition}\label{defn:mfailuresCode}
An $[n,k,d_{min}]_2$ protection code $\C$ is a $k$ dimensional subspace of
the space $\F_2^{n}$ that is able to  protect all network failures
up to $ d_{min}-1 $.
\end{definition}

\smallskip

In general the \emph{network protection code} (NPC), which protects against
multiple path failures, can be defined by a generator matrix $G$
known for every sender and receiver. Also, there exists a parity check matrix $H$ corresponds to $G$ such that $GH^T=0$.
 We will restrict ourselves in this work
to NPC that are generated by a given systematic generator matrix $G$ over $\F_2$.

Without loss of generality, the protection matrix among all sources is given by

\begin{eqnarray}
\begin{array}{c||cccccccc}
&L_1&L_2&\cdots&L_k &L_{k+1}&L_{k+2}&\ldots&L_{n}\\
 \hline \hline
s_1\!\!&x_1&0&\cdots&0&p_{11}x_{1}&p_{12}x_{1}&\!\!\ldots&\!\!p_{1m}x_{1}\\
s_2\!\!&0&x_2&\cdots&0&p_{21}x_{2}&p_{22}x_{2}&\!\!\ldots&\!\!p_{2m}x_{2}\\
\vdots\!\!&\vdots&\vdots&\cdots&\vdots&\vdots&\!\!\vdots&\!\!\vdots\\
s_{k}\!\!&0&0&\cdots&x_k&p_{k1}x_{k}&p_{k2}x_{k}&\!\!\ldots&\!\!p_{km}x_{k}\\
\hline \hline \!\!&x_1&x_2&\ldots&x_k&y_{k+1}& y_{k+2}&\!\!\ldots&\!\!y_{n}\\
\end{array}
\end{eqnarray}

We ensure that $k=n-m$ paths have full capacity and they carry the
plain data $x_1,x_2,\ldots,x_k$. Also, all other $m$ paths have full capacity, in which
they carry the encoded data $y_{k+1},y_{k+2},\ldots,y_{n}$. In addition, the $m$
links are not fixed, and they are chosen alternatively between the
$n$ links.


 \noindent \textbf{Encoding Process.} The network encoding
processes at the set of senders are performed in a similar manner as
in Section~\ref{sec:singlefailure}. Every source $s_i$ has a copy of
the  systematic matrix $G$ and it will prepare a packet along with its ID in
two different cases. First, if the source $s_i$ will send only its
own data $x_i$ with a full link capacity, then

\begin{eqnarray}
packet_{s_i} = (ID_{s_i}, x_{i}, t_i^\delta).
\end{eqnarray}

Second, if the source $s_j$ will send an encoded data in its packet,
then

\begin{eqnarray}
packet_{s_j} = (ID_{s_j},  \sum_{\ell=1, \ell\neq
j}^k p_{\ell j}x_\ell, t_j^\delta),
\end{eqnarray}
where $p_{\ell j} \in \F_2$.

\noindent \textbf{Recovery Process.} The recovery process is done as
follows. The  $packet_{s_i}$ arrives at a receiver $r_i$ in time
slots, hence every packet from a source $s_i$ has a round time
$t_i^\delta$. In this case, time synchronization is needed to guarantee
the reception of the correct data. The receiver $r_i$ at time slot
$n$ will detect the signal in the link $L_i$. If the link $L_i$
failed, then $r_i$ will send a query to other receivers in $R
\backslash \{r_i\}$ asking for their received data. Assume there are
$t $ path failures. Then we have three
cases:
\begin{compactenum}
\item All $t$ link failures have occurred in links that do not carry encoded
packets, i.e., $packet_{s_i} = (ID_{s_i}, x_{i}, t_i^\delta)$. In
this case, one receiver that carries an encoded packet, e.g., $r_j$, can send
$n-t-1$ queries to the other receivers with active links asking for
their received data. After this
process, the receiver $r_j$ is able to decode all messages. \nix{ and will
send individual messages to all receivers with link failures to pass
their correct data.}

\item
All $t$ link failures have occurred in links that  carry encoded
packets, i.e., $packet_{s_j} = (ID_{s_j},
\sum_{\ell=1, \ell\neq j}^k x_\ell, t_j^\delta)$. In this case no recovery operations are needed.

\item
All $t$ link failures have occurred in arbitrary links. This case is
a combination of the previous two cases and the recover process is
done in a similar way. Only the lost data on the working paths needs to be recovered.
\end{compactenum}

\goodbreak

Our future work will include practical implementation aspects of the proposed model as shown in the case of adding extra paths~\cite{kamal07b}. The proposed
network protection scheme using distributed capacity and coding is able to recover up
to  $t \leq d_{min}-1 $ link failures (as defined in Definition \ref{defn:mfailuresCode})
among $n$ paths and it has the following advantages:
\begin{compactenum}[i)]
\item
$k=n-m$ links have full capacity and their sender nodes have the same
transmission rate.
\item
The $m$ links that carry encoded data are dynamic (distributed) among all $n$
links. Therefore, no single link $L_i$ will suffer from reduced
capacity.

\item
The encoding process is simple once every sender  $s_i$ knows the
NPC. Hence $s_i$ maintains a round time $t_\ell^\delta$ for each sent
$packet_{s_i}$.
\item The recovery from link failures is done in a dynamic and simple
way at one receiver. \nix{Only one receiver node needs to perform the decoding process
and it passes the data to other receivers that suffer from link failures.}
\end{compactenum}

\section{Analysis}\label{sec:analysis}

We shall provide theoretical analysis regarding the proposed  network
protection codes. One can easily compute the number of paths needed
to carry encoded messages to protect against $t$ link failures, and
will obtain the average network capacity. The main idea behind NPC is to
simplify the encoding operations at the sources and the decoding
operations at the receivers. The following lemma demonstrates the
average normalized capacity of the proposed network model $\N$.

\begin{lemma}\label{lem:capacitymulitplelinks}
Let $\C$ be a protection code with parameters $ [n,n-m,d_{min}]$ over $\F_2$. Assume  $n$ and $m$ be the number of sources (receivers) and number of
connections carrying encoded packets, respectively, the average capacity of the
network $\N$ is given by
\begin{eqnarray}
(n- m )/n.
\end{eqnarray}
\end{lemma}
\begin{proof}
We have m protection paths that carry encoded data. Hence there are $n-m$ working paths that carry plain data. The result is a direct consequence by applying the normalized capacity definition.
\end{proof}

\goodbreak

\begin{lemma}
In the  network protection model $\N$, in order to protect $t$
network disjoint link failures, the minimum distance of the protection code must be at least $t+1$.
\end{lemma}
\begin{proof}
We can assume that the network link failures can occur at any
arbitrary paths. The proof comes from the fact that the protection code can detect $t$ failures.
\end{proof}

The previous lemma ensures that the maximum number of failures that can be recovered by $\C$ is $d_{min}-1$.

For example one can use the Hamming codes with parameters $[2^\mu,2^\mu-\mu-1,3]_2$, for some positive integer $\mu$, to recover from two failures, see ~\cite{huffman03,macwilliams77} for notation. One can also puncture these codes to reach the required length, i.e., number of connections.  $[7,4,3]_2$, $[15,11,3]_2$, and $[63,57,3]_2$ are examples of  Hamming codes that protect against two link failures.  Another example is the BCH codes with arbitrary design distance. $[15,11,3]_2$, $[31,26,3]_2$ and $[63,56,3]_2$ are examples of BCH codes that protect one and two link failures. Also,  $[15,8,5]_2$, $[31,21,5]_2$ and $[48,36,5]_2$ are examples of BCH codes that protect against four link failures~\cite{huffman03,macwilliams77}.

\section{Conclusion}\label{sec:conclusion}

We studied a model for recovering from network link failures using
network coding and reduced capacity. We defined the concept of
\emph{network protection codes} to protect against arbitrary $t$ link
failures. We showed that the encoding and decoding processes of the proposed scheme are simple and can be done in a dynamic way at any arbitrary
senders and receivers in an overlay layer on optical mesh networks.
Our future work will include  tables of best known protection codes and a comparison between protection against
link failures using reduced capacity and using  extra paths.

\scriptsize
\bibliographystyle{plain}

\end{document}